\documentclass[epj,final]{svjour}
\usepackage{graphics}
\usepackage{epsfig}
\usepackage{amsmath}

\begin{document}
\title{\protect$\eta'\text{q}\bar{\text{q}}$ and 
       \protect$\eta'\text{NN}$ vertex suppression in  effective theories} 
\author{D.~Lehmann\inst{1}}
\institute{Institut f{\"u}r Theoretische Physik~III,
  Universit{\"a}t Erlangen-N{\"u}rnberg,\protect\\
  Staudtstra{\ss}e~7, D-91058 Erlangen, Germany,
  \email{dirkleh@theorie3.physik.uni-erlangen.de}}

\abstract{
  In an effective theory containing only quark degrees of freedom,
  such as the extended Nambu--Jona--Lasinio model, the influence
  of the axial anomaly can be incorporated by a self-interaction
  of the 't~Hooft determinant type.
  I will show that despite the theshold problem related to the 
  $\eta'$ meson this leads to a significant suppression of the
  $\eta'$ coupling $g_{\eta' \text{q}\overline{\text{q}}}$ to
  dynamical quarks which suggests a suppression of
  the $\eta'$NN vertex as compared to the $\eta$NN vertex.
  }

\maketitle

The recent experimental efforts in $\eta'$ electro- and
photo-production in facilities like ELSA, JLAB, DA$\Phi$NE,
and GRAAL as well the recent close to
threshold $\text{pp}\to\text{pp}\eta'$ production results of
COSY-11 \cite{COSY} have increased the interest in $\eta'$
interaction properties. So far, only little is known about the
$\eta'$NN vertex, both theoretically and experimentally.
In effective meson theories the vertex is \emph{para\-metrized}
through a single coupling constant
$g_{\eta'\text{NN}}$ ($1.9 \le g_{\eta'\text{NN}} \le 6.2$, see e.g.\ 
Ref.~\cite{Zhang}) rather than predicted from an underlying dynamics.

As is well-known, the $\eta'$ meson differs significantly in
its mass from the pseudo-scalar octet mesons.
This comes about, since the $\eta'$ is predominantly a
flavor-singlet, and the axial anomaly of QCD breaks the
axial U(1) symmetry,
preventing the $\eta'$ from being a pseudo-Goldstone boson unlike
its octet counterparts.
It is therefore very likely that the axial anomaly has also a
significant influence on the interaction properties of the
$\eta'$ particle, e.g.\ on the  $\eta'\text{NN}$ vertex.

Since a direct calculation of the coupling in QCD is presently not feasable,
one has to resort to specific models. 
A model that has proven both, tractable and phenomenologically 
successful in describing properties of
low-lying mesons is the three flavor extended
Nambu--Jona--Lasinio (ENJL) model 
(see e.g.\ Refs.~\cite{Klimt,Klevansky,Weise} for a review).
It not only exhibits the chiral symmetry breaking aspects of the
pseudo-scalar octet mesons but also allows a direct inclusion
of the U(1)$_{\mathrm{A}}$ breaking effects of the axial anomaly. 
The later is taken into account by an additional local 
six-fermion self-interaction of the 't~Hooft determinant type \cite{tHooft}
and turned out to be of crucial importance for the $\eta$ and
$\eta'$ mesons.
While the ENJL model has been successfully applied to the $\eta$ meson,
see e.g.\ Ref.~\cite{Takizawa},  it has rarely been applied to flavor 
singlet $\eta'$, since the lack of confinement,
one of the major shortcomings of the model, will cause the $\eta'$
meson to lie above the $\text{q}\overline{\text{q}}$ threshold
when realistic parameters are chosen.
This results in unphysical decay modes
$\eta'\to\text{q}\overline{\text{q}}$.
Thus, care is needed to identify the pole
on the second Riemann sheet reasonably well, see e.g.\
Refs.\ \cite{Meissner,Lemmer}.

\begin{figure}[ht]
  \centering
  \epsfig{file=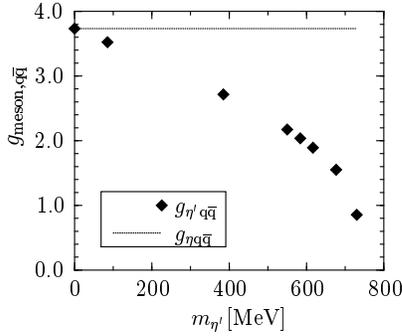, width=0.3\textwidth}
  \caption{Suppression of the $\eta'\text{q}\overline{\text{q}}$
   coupling (diamonds) in dependence of the $\eta'$ mass in 
   the chirally symmetric ENJL model. The line indicates the
   constan value for the octet mesons.
    (Data taken from Ref.~\cite{PhD}.)}
                \label{Fig:1}
\end{figure}

In Ref.~\cite{Hatsuda} it has been shown that due to the 't~Hooft 
interaction the $\eta'$-quark coupling
$g_{\eta'\text{q}\overline{\text{q}}}$ is significantly suppressed 
compared to the other pseudo-scalar mesons. However, 
no particular attention has been paid to the theshold problem. 
When realistic parameters are chosen, quite some imaginary part is 
obtained in $g_{\eta'\text{q}\overline{\text{q}}}$ which reflects 
the unphysical decay mode $\eta'\to \text{q}\overline{\text{q}}$.
The main result of this note is to demonstrate that the
observed suppression is indeed physically sensible and not
just a threshold artifact.
For that purpose I investigated \cite{PhD} a chirally symmetric version of
the model with a dynamical quark mass of $407\,\text{MeV}$.
Here a wide $\eta'$ mass range of about $800\,\text{MeV}$ can be
covered \emph{without} encountering threshold effects.
The results shown in Fig.~\ref{Fig:1} conclusively support the
singlet ($g_{\eta'\text{q}\overline{\text{q}}}$) suppression
compared to the (constant) octet coupling. A threshold artifact thus
seems to be ruled out.

\begin{figure}[!ht]
  \centering
  \epsfig{file=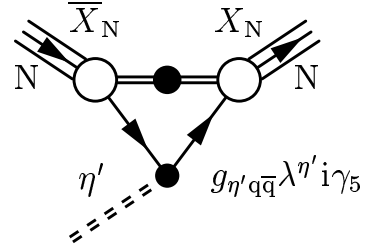, width=0.3\textwidth}
  \caption{Leading contribution to the \protect$\eta'\text{NN}$ 
    vertex in an effective meson-quark theory. The double line
    denotes the scalar diquark propagator, while
    \protect$X_{\text{N}}$ and \protect$\overline{X}_{\text{N}}$
    are nucleon-quark-diquark vertices.}  
                                                       \label{Fig:2} 
\end{figure}

We have to be aware that the observed $\eta'$ suppression
\emph{a priori\/} depends crucially on our model choice.
It is therefore essential to discuss in how far this is believed
to hold in QCD as well.
To this end, note that the origin of the
$g_{\eta'\text{q}\overline{\text{q}}}$ suppression in the ENJL model
is a destructive interference between the attractive part of the
standard four-fermion interaction and the
(in the flavor-singlet channel) \emph{repulsive} 't~Hooft
interaction.
In the octet sector in contrast, the 't~Hooft interaction acts
attractive, conspiring with the four-fermion interaction to fulfill Goldstone's
theorem in the chiral limit.\footnote{%
  Of course, the axial U(1)-breaking also induces a singlet-octet
  mixing term,
  which in the absence of flavor-symmetry complicates the argumentation
  slightly without spoiling it.
}
A destructive interference mechanism, however, strongly supports the
\emph{model independence} of my conclusions, since in any
description of the pseudo-scalar meson sector the axial anomaly
will contribute repulsively in the singlet ($\approx \eta'$) channel
to account for the $\eta'$ mass, while an attractive part is most
likely present and responsible for the binding of the flavor-octet
pseudo-Goldstone bosons and essentially compensates the
large constituent quark masses.
The suppression is important for the physically observable
$\eta'\text{NN}$ vertex, which may be calculated within the ENJL
model to leading order as shown in Fig.~\ref{Fig:2}. The nucleon
can be described by solving the relativistic Faddeev equations in the
quark-diquark picture with a quark-exchange interaction, following
the approaches of
Refs.~\cite{Ishii,Huang}.
An even simpler approximation consists in the static approximation
\cite{PhD,Buck}, i.e.\ assuming the exchanged quark to
be infinitely heavy. Since the nucleon-quark-diquark vertices are
not sensitive neither to the 't~Hooft interaction nor to the meson
channel, the ratio of $\eta'\text{NN}$ and $\eta\text{NN}$ vertices
is in leading order determined by the ratio of meson-quark couplings,
\begin{equation}
  \frac{ \mathcal{V}_{\eta'\text{NN}}}{\mathcal{V}_{\eta\text{NN}}}
  =
  \frac{g_{\eta'\text{q}\overline{\text{q}}}}
  {g_{\eta\text{q}\overline{\text{q}}}} \: 
  \frac{   \sqrt{\frac{2}{3}} \cos \theta_{\eta'}
    + \sqrt{\frac{1}{3}} \sin\theta_{\eta'} }
  { - \sqrt{\frac{2}{3}} \sin \theta_{\eta}
    + \sqrt{\frac{1}{3}} \cos\theta_{\eta}  }
  \approx 0.30 + 0.22\,\mathrm{i} \enspace.
\end{equation}
A comparison of $\eta'$ vs.\  $\pi^{0}$ productions is
of similar magnitude,
\begin{equation}
  \frac{ \mathcal{V}_{\eta'\text{NN}} }{ \mathcal{V}_{\pi\text{NN}} }
  \approx 0.21 - 0.15\,\mathrm{i}\enspace.
\end{equation}
A suppression of the $\eta'\mathrm{NN}$ vertex conforms with 
claims in the literature, e.g.\ in the  
the context of the EMC effect ("spin crisis"), see Refs.\
\cite{Veneziano89,ShoreVeneziano,Efremov,Hatsuda} or at finite density
\cite{Kharzeev}.

\begin{acknowledgement}
  Its a pleasure to thank F.~Lenz and  K.~Yazaki for fruitful
  discussions and U.-G.~Meissner and R.~H.~Lemmer for helpful hints.
  This work was supported by the Bundesministerium f{\"u}r
  Bildung und Forschung (BMBF) under grant 06~ER~809.
\end{acknowledgement}

%
%
%

\end{document}